\documentclass[prd,nofootinbib,floatfix,10 pt,superscriptaddress,eqsecnum,twocolumn]{revtex4-1}

\usepackage{graphicx,wrapfig,float,slashed,subcaption}
\usepackage{amsmath,amssymb,epsfig,graphicx}
\usepackage{multirow}
\usepackage{ragged2e}
\usepackage{array}

\usepackage{cleveref}
\newcommand {\tanb}{\tan\beta}

\begin{document}

\title{Leptophilic neutral Higgs bosons in two Higgs doublet model at a linear collider}

\author{Majid Hashemi}
\affiliation{Physics Department, College of Sciences,\\
Shiraz University, Shiraz, 71946-84795, Iran}
\email{hashemi\_mj@shirazu.ac.ir}

\begin{abstract}
This paper addresses the question of observability of neutral Higgs bosons through the leptonic decay in a two Higgs doublet model (2HDM). Both scalar and pseudoscalar Higgs bosons ($H,~A$) are considered. The model is set to type IV to enhance the leptonic decay. In such a scenario, a signal production process like $e^+e^- \to A^0H^0 \to \tau\tau \mu\mu$ or $\mu\mu\tau\tau$ would provide a clear signal on top of the background in a di-muon invariant mass distribution far from the $Z$ boson pole mass. The analysis is based on a $\tau$-id algorithm which preselects events if they have two $\tau$ jets by requiring a hadronic $\tau$ decay. Several benchmark points are defined for the search, requiring a linear collider operating at $\sqrt{s} =$ 0.5 and 1 TeV. It is shown that the signal can be observed on top of the background in all benchmark points at an integrated luminosity of 1000 $fb^{-1}$.
\end{abstract}


\maketitle

\section{Introduction}
\label{Introduction}
The Standard Model of Particle Physics (SM) has been tested successfully in a large number of experiments. The Higgs boson introduced through a single SU(2) doublet \cite{Englert1,Higgs1,Higgs2,Kibble1,Higgs3,Kibble2} has been observed at LHC \cite{HiggsObservationCMS,HiggsObservationATLAS} with a mass near 125 GeV. The ongoing analyses at LHC confirm that the observed boson has the same properties expected for a Standard Model Higgs. However, it is still an open question whether the observed particle belongs to a single SU(2) doublet or a richer framework such as a two Higgs doublet model \cite{2hdm1,2hdm2,2hdm3}.

The two Higgs doublet models are well motivated beyond SM from different aspects. They provide the basis for building weakly coupled theories such as the Minimal Supersymmetric extention of Standard Model (MSSM) \cite{MSSM1,MSSM2,MSSM3} as well as the strongly coupled Composite Higgs model \cite{2hdm4_CompositeHiggs}. There are four types of 2HDMs which are designed to accomodate different possible scenarios of Higgs-fermion couplings. The ratio of vacuum expectation values of the two Higgs doublets ($\tan\beta=v_2/v_1$) is a key tool in all 2HDM types \cite{tanbsignificance}.

In general, 2HDMs (and MSSM) involve more than a single physical Higgs boson due to the second Higgs doublet which is added to the SM resulting in a larger degrees of freedom available to the model after giving masses to the gauge bosons. Taking the lighest scalar, $h$, as the SM-like Higgs boson, there are two more neutral Higgs bosons, $H,~A$, and two charged bosons, $H^{\pm}$. Contrary to the MSSM which involves almost degenerate heavy Higgs boson states, the 2HDM allows for different Higgs boson masses, thus providing a broader parameter space available for study. The theory and phenomenology of 2HDMs has been discussed in detail in \cite{2hdm_TheoryPheno}. In addition to direct searches, the 2HDM is also a key tool in flavor Physics for bringing theory predictions close to experimental observations by including processes which involve 2HDM Higgs bosons \cite{FMahmoudi}.

In this work, different benchmark points are defined in the Higgs boson mass spectrum space and a search is performed at a linear collider. The signal process is taken as a joint $AH$ production through $e^+e^- \to AH \to \mu\mu\tau\tau$ or $\tau\tau\mu\mu$. It is expected that a leptonic decay of the neutral Higgs ($H$ or $A$) would provide a clear signal due to the reasonable lepton reconstruction efficiency at linear colliders. For such a process, the 2HDM type IV is chosen to enhance the leptonic Higgs decay at high $\tanb$. There has been a recent work searching for the same signal at LHC \cite{kanemura}. In this analysis, it is shown that signals of both $H$ and $A$ bosons are observable at a future linear collider. Details of the model and benchmark points are presented in the next section.

\section{The Higgs sector of 2HDM}
The Higgs sector of 2HDM involves neutral and charged Higgs couplings with fermions (leptons and quarks). The Lagrangian for neutral Higgs-fermion couplings as introduced in \cite{2hdm_HiggsSector1} is written in Eq. \ref{lag}, where $\kappa^f=\sqrt{2}\frac{m_f}{v}$ for any fermion type $f$. The four types of interactions (2HDM types) are defined as in Tab. \ref{types} \cite{Barger_2hdmTypes}. The type III is sometimes called ``flipped'' or ``type Y'' and the type IV is also known as ``lepton-specific'' or ``type X''. The collider phenomenolgy of 2HDM depends on the model type which determines which kind of Higgs-fermion interactions are more important for a given $\tanb$ \cite{2hdm_HiggsSector2}.

In order to respect SM observations, the lightest Higgs boson, $h$, is taken to be SM-like by setting $s_{\beta-\alpha}=1$. This ensures that the $s_{\beta-\alpha}$ factor in the lightest Higgs-gauge coupling is set to unity while the heavier Higgs, $H$, decouples from gauge bosons \cite{2hdm_TheoryPheno}. On the other hand, terms containing $\rho$ are eliminated to remove the $\tanb$ dependence of the SM-like Higgs-fermion interactions. Therefore all SM-like Higgs interactions with fermions and gauge bosons are equal to their SM corresponding values.
\begin{widetext}
\begin{align}
\begin{split}
-\mathcal{L}=&\frac{1}{\sqrt{2}} \bar{D}\left\{\kappa^D s_{\beta-\alpha}+\rho^D c_{\beta-\alpha} \right\}Dh
+\frac{1}{\sqrt{2}} \bar{D}\left\{\kappa^D c_{\beta-\alpha}-\rho^D s_{\beta-\alpha} \right\}DH+\frac{i}{\sqrt{2}}\bar{D}\gamma_5\rho^D DA\\
&\frac{1}{\sqrt{2}} \bar{U}\left\{\kappa^U s_{\beta-\alpha}+\rho^U c_{\beta-\alpha} \right\}Uh
+\frac{1}{\sqrt{2}} \bar{U}\left\{\kappa^U c_{\beta-\alpha}-\rho^U s_{\beta-\alpha} \right\}UH-\frac{i}{\sqrt{2}}\bar{U}\gamma_5\rho^U UA\\
&\frac{1}{\sqrt{2}} \bar{L}\left\{\kappa^L s_{\beta-\alpha}+\rho^L c_{\beta-\alpha} \right\}Lh
+\frac{1}{\sqrt{2}} \bar{L}\left\{\kappa^L c_{\beta-\alpha}-\rho^L s_{\beta-\alpha} \right\}LH+\frac{i}{\sqrt{2}}\bar{L}\gamma_5\rho^L LA\\
\end{split}
\label{lag}
\end{align}
\end{widetext}
\begin{table}
\centering
\begin{tabular}{|c|c|c|c|c|}
\hline
\multicolumn{5}{|c|}{Type}\\
& I & II& III&IV\\
\hline
$\rho^D$ & $\kappa^D \cot\beta$ &$-\kappa^D \tan\beta$ &$-\kappa^D \tan\beta$ &$\kappa^D \cot\beta$  \\
\hline
$\rho^U$ & $\kappa^U \cot\beta$ &$\kappa^U \cot\beta$ &$\kappa^U \cot\beta$ &$\kappa^U \cot\beta$  \\
\hline
$\rho^L$ & $\kappa^L \cot\beta$ &$-\kappa^L \tan\beta$ &$\kappa^L \cot\beta$ &$-\kappa^L \tan\beta$  \\
\hline
\end{tabular}
\caption{Different types of 2HDM in terms of the Higgs boson couplings with $U$(up-type quarks), $D$(down-type quarks) and $L$(leptons).\label{types}}
\end{table}

Incorporating Flavor Physics data results, the type II and III will receive a strong lower limit on the chaged Higgs mass at 480 GeV \cite{Misiak}. The type I is interesting for low $\tanb$ as all couplings in the neutral Higgs sector are proportional to $\cot\beta$. A study of neutral Higgs decays in this type at LHC shows that a pseudo-scalar Higgs production followed by the decay $A\to ZH$ can be observed with $H$ decaying to $b\bar{b}$ or $WW$ \cite{TypeI_LHC}.

On the other hand, type IV provides a Higgs-lepton coupling which enhances as $\tanb$. Therefore at high $\tanb$ all neutral Higgs, $H$, couplings are suppressed except for the leptonic decays. Such a ``$leptophilic$'' Higgs can be observed in a di-lepton invariant mass distribution on top the background. The lepton in this case is either $\tau$ or $\mu$. The branching ratio of Higgs decay to $\tau$ is higher because of the larger mass. However, identification of such decays requires a hadronic $\tau$-id which reconstructs the hadronic part of the decay. On the other hand, the muonic decay is easy to reconstruct and observe, as the muon trajectory is well identified at a linear collider. The Higgs branching ratio of decay to muons is 3 permil in the best case.
\section{Benchmark points and cross sections}
\begin{figure}[]
\centering
\begin{subfigure}{0.45\textwidth}
  \includegraphics[width=\textwidth]{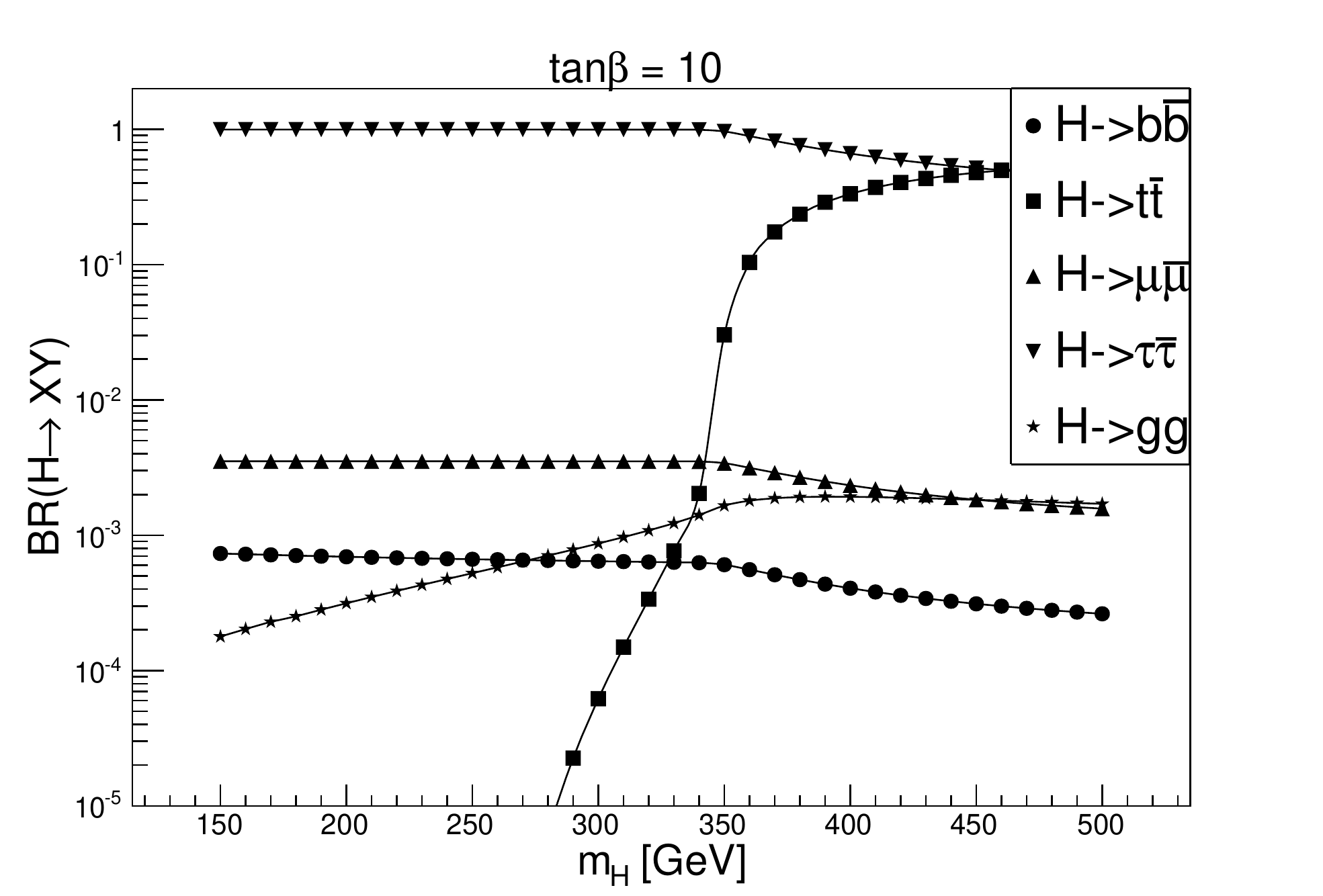}
  \caption{}
  \label{DecayH0}
\end{subfigure}%
\,\,
\begin{subfigure}{0.45\textwidth}
  \includegraphics[width=\textwidth]{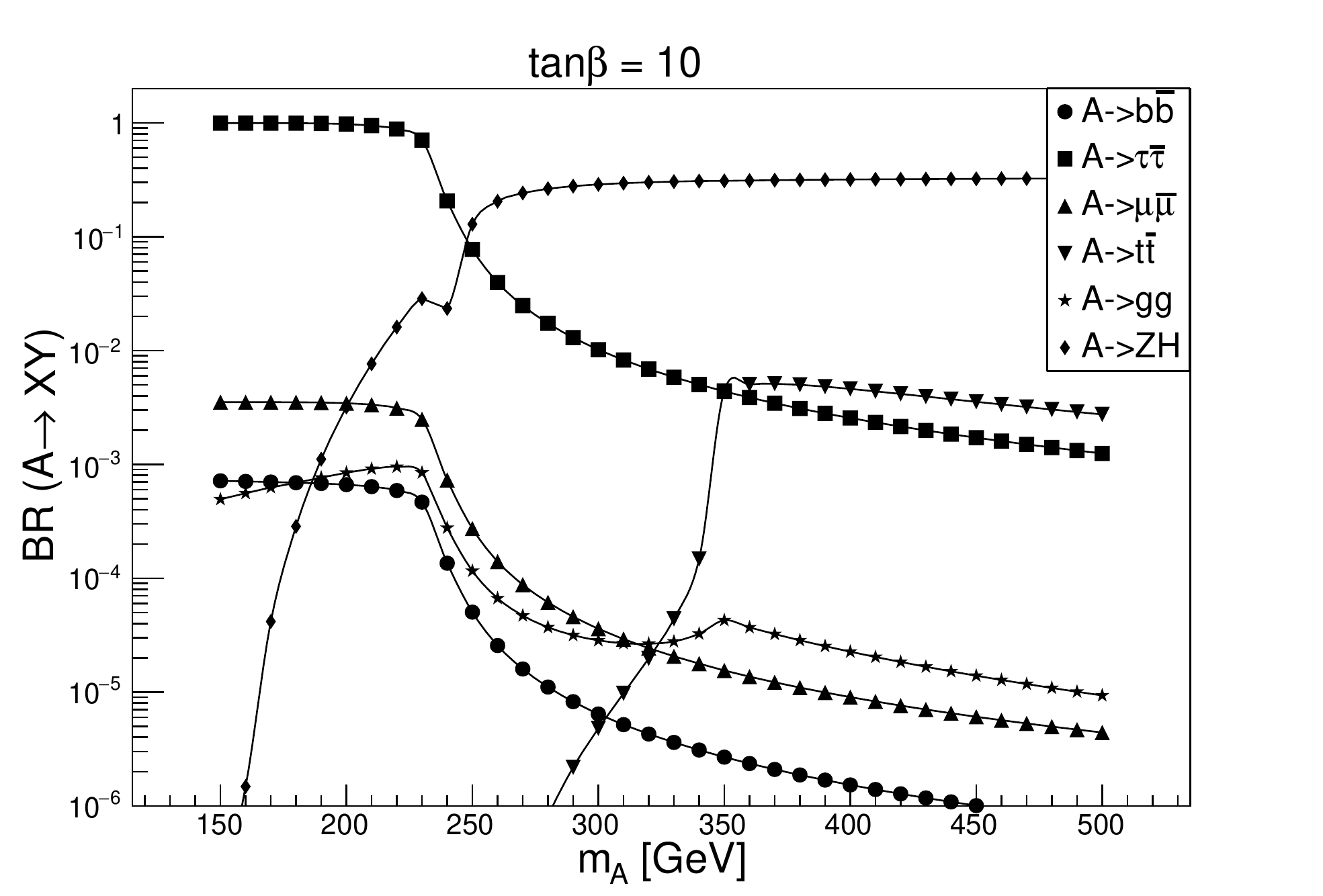}
  \caption{}
  \label{DecayA0}
\end{subfigure}%
\caption{The $A$ and $H$ branching ratio of decays. On the right plot $m_H=150$ GeV.}
\label{Hdecay}
\end{figure}

In order to select working points, it is better to plot branching ratio of $H$ and $A$ decays in a 2HDM type IV as shown in \cref{DecayH0} and \cref{DecayA0}. The scalar $H$ decays to $\tau$ or $\mu$ for masses below the threshold of top pair production, i.e., $m_H\simeq 350$ GeV. A heavier Higgs would prefer to decay to $t\bar{t}$. Therefore the analysis is limited to $m_H$ below the top pair threshold. The pseudo-scalar Higgs A also decays to $\tau$ and $\mu$ until $A \to ZH$ starts to be kinematically possible.

The scenario followed in this paper assumes that the pseudo-scalar Higgs, $A$, is heavier than $H$. For a given $m_H$, if $m_A$ increases to regions where $m_A>m_H+m_Z$, the type independent decay $A\to ZH$ occurs resulting in a suppression of leptonic decays. Therefore all benchmark points are chosen not to allow such a decay to overhelm the leptonic decays. The mass difference $m_A-m_H$ is thus adjusted to be less than $m_A$, leaving no phase space for the decay products. The attempt is to search for moderate and high masses of neutral Higgs bosons. Therefore the analysis covers linear colliders with $\sqrt{s}=0.5$ TeV and 1 TeV. The selected benchmark points (BP's) are presented in Tab. \ref{bps}. They cover $H$ masses in the range 150-300 GeV. The charged Higgs mass is set to $m_A$ to reduce $\Delta\rho$ \cite{drho}.
The first four points are studied at $\sqrt{s}=0.5$ TeV and the rest at $\sqrt{s}=1$ TeV. The BP4 and BP5 are in fact the same points. As will be shown, this point is not observable at 0.5 TeV but has a signal exceeding 5$\sigma$ at 1 TeV.
\begin{table}[h]
\centering
\begin{tabular}{|c|c|c|c|c|c|c|}
\hline
\multicolumn{7}{|c|}{$\sqrt{s}=500$ GeV}\\
BP & $m_h$ & $m_H$ & $m_A$ & $m_{H^{\pm}}$ & $s_{\beta-\alpha}$ & $\tanb$\\
\hline
1 & \multirow{4}{*}{125} & 150 & 150 & 150 & \multirow{4}{*}{1} & \multirow{4}{*}{10} \\ \cline{1-1} \cline{3-5}
2 &  & 150 & 200 & 200 &  &  \\\cline{1-1} \cline{3-5}
3 &  & 200 & 200 & 200 &  &  \\\cline{1-1} \cline{3-5}
4 &  & 200 & 250 & 250 &  &  \\
\hline
\multicolumn{7}{|c|}{$\sqrt{s}=1000$ GeV}\\
BP & $m_h$ & $m_H$ & $m_A$ & $m_{H^{\pm}}$ & $s_{\beta-\alpha}$ & $\tanb$\\
\hline
5 & \multirow{4}{*}{125} & 200 & 250 & 250 & \multirow{4}{*}{1} & \multirow{4}{*}{10} \\ \cline{1-1} \cline{3-5}
6 &  & 250 & 300 & 300 &  &  \\\cline{1-1} \cline{3-5}
7 &  & 300 & 300 & 300 &  &  \\\cline{1-1} \cline{3-5}
8 &  & 300 & 350 & 350 &  &  \\
\hline
\end{tabular}
\caption{The benchmark points selected for linear colliders operating at $\sqrt{s}=500$ and 1000 GeV. The charged Higgs mass is set to the heaviest boson mass. The BP5 is the same point as BP4 but searched for at a different center of mass energy.}
\label{bps}
\end{table}

\begin{figure}[]
\centering
\begin{subfigure}{0.45\textwidth}
  \includegraphics[width=\textwidth]{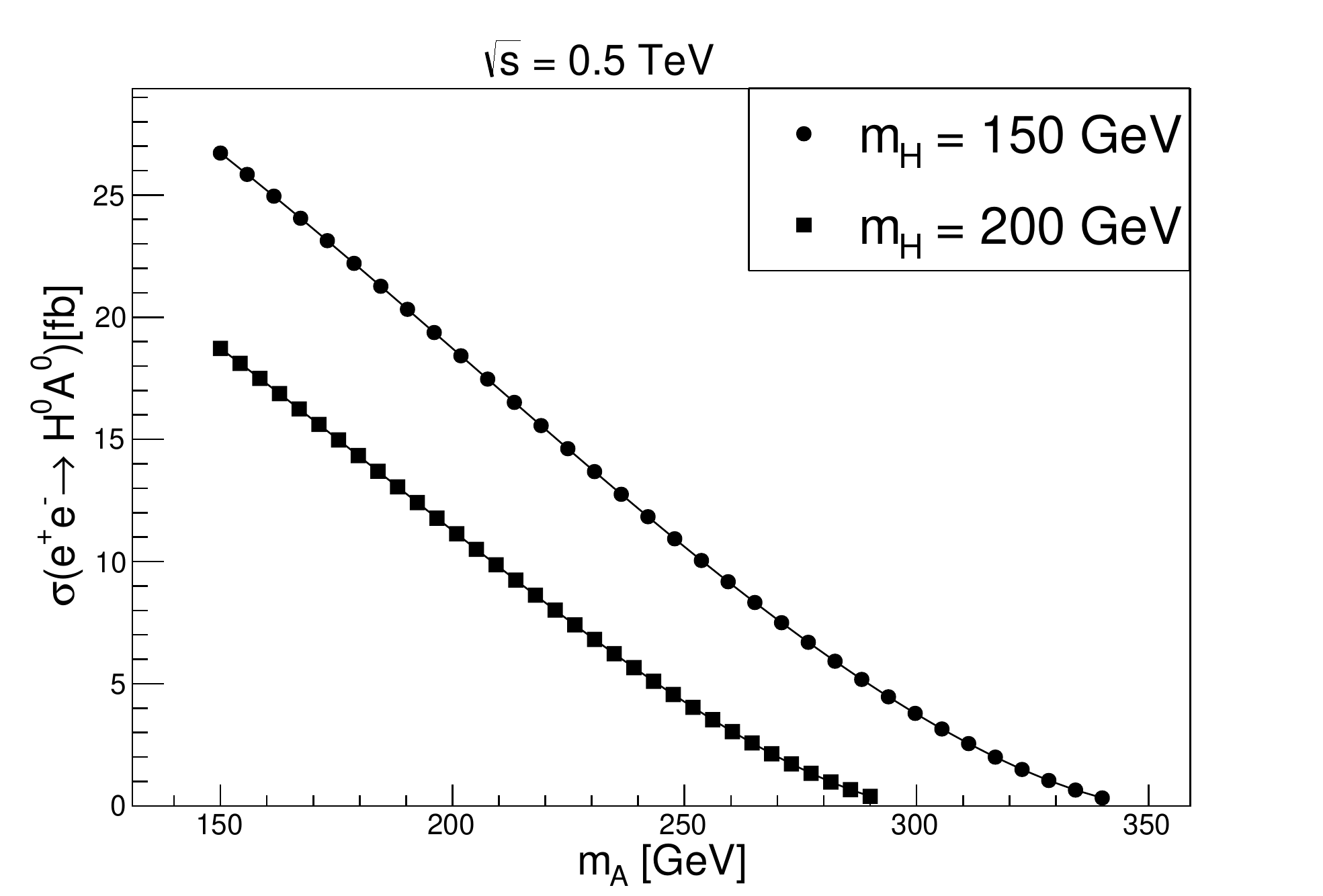}
  \caption{}
  \label{sigma05}
\end{subfigure}%
\,\,
\begin{subfigure}{0.45\textwidth}
  \includegraphics[width=\textwidth]{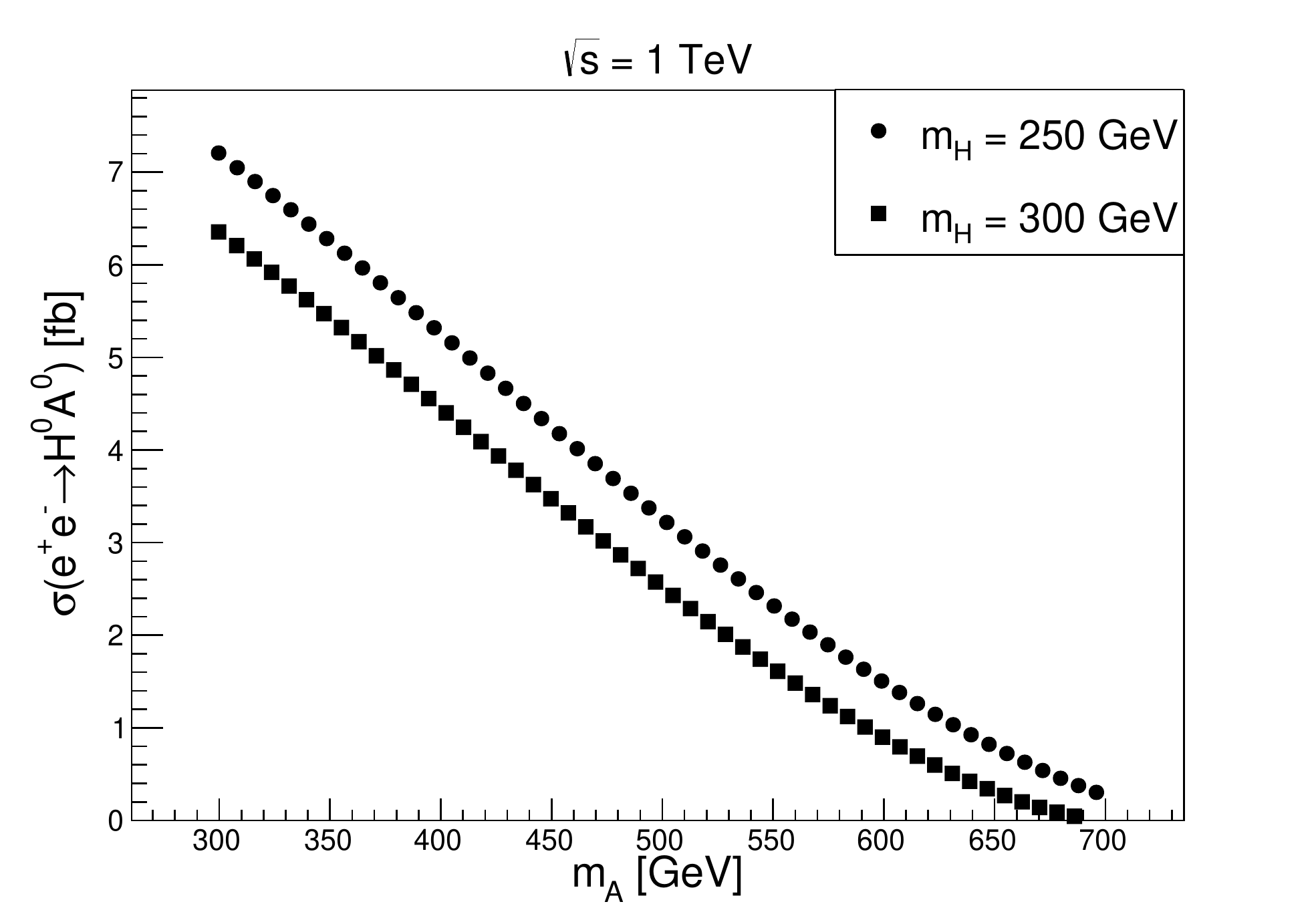}
  \caption{}
  \label{sigma1}
\end{subfigure}%
\caption{The signal cross section at a linear collider operating at $\sqrt{s}=0.5$ TeV (a), and $\sqrt{s}=1$ TeV (b) as a function of the two neutral Higgs boson masses.}
\label{}
\end{figure}

The signal cross sections are shown in Fig. \ref{sigma05} and \ref{sigma1}. The signal process, $e^+e^- \to AH$, depends on $s_{\beta-\alpha}$ which is set to 1. Therefore increasing $\tanb$ has no effect on the production cross section. On the other hand, the leptonic decays of $A$ and $H$ are dominated by $\tau$ leading to BR$(H \to \tau\tau)\simeq 1$ at $\tanb\simeq 10$. Again increasing $\tanb$ to higher values has no sizable effect on the signal rate through the Higgs boson decays. The signal cross section is sensitive to the mass difference $\delta=m_A-m_H$ and is reduced when $\delta$ decreases. Therefore a high cross section prefers a large mass splitting which is avoided by the requirement of leptonic decay enhancement and $A \to ZH$ suppression.

The main SM background processes are $WW$, $ZZ$, $Z/\gamma^*$ and $t\bar{t}$. Their cross sections are listed in \cref{cs} for the two center of mass energies.
\begin{table}[h]
\centering
\begin{tabular}{|c|c|c|c|c|}
\hline
\multicolumn{5}{|c|}{$\sqrt{s}=0.5$ TeV}\\
 & WW & ZZ & $Z/\gamma^*$ & $t\bar{t}$\\
\hline
$\sigma [pb]$ & 7.83 & 0.58 & 16.67 & 0.59 \\
\hline
\multicolumn{5}{|c|}{$\sqrt{s}=1$ TeV}\\
 & WW & ZZ & $Z/\gamma^*$ & $t\bar{t}$\\
\hline
$\sigma [pb]$ & 3.19 & 0.2 & 4.3 & 0.21 \\
\hline
\end{tabular}
\caption{The background cross sections at $\sqrt{s}=0.5$ and 1 TeV.}
\label{cs}
\end{table}
\section{Signal selection and analysis}
Signal and background events are generated using {\tt PYTHIA 8} \cite{pythia}. Jets are reconstructed using {\tt FASTJET 2.8} \cite{fastjet1,fastjet2} based on anti-$k_T$ algorithm with a cone size of 0.4. The jet energy is smeared according to energy resolution $\sigma/E=3.5\%$ predicted for CLIC \cite{cliccdr}. Only jets which pass a kinematic threshold as in \cref{kin} are selected.
\begin{equation}
E_{T}^{\textnormal{jet}}>10~\textnormal{GeV},~ |\eta|<1.5
\label{kin}
\end{equation}

These jets make the seed for the $\tau$ identification. The $\tau$ jets are characterized as isolated narrow jets because they consist of few charged particles (pions) flying collinearly. A normal jet from QCD interactions accomodates a large number of tracks. The $\tau$ identification algorithm starts with the isolation requirement which verifies that there is no track with $p_T>1$ GeV in an annulus defined by $0.07<\Delta R<0.1$ around the jet hottest track. The next key feature is the number of tracks inside the signal cone defined as a narrow cone ($\Delta R=0.07$) around the hottest track. Since $\tau$ jets predominantly decay to one or three charged pions in their hadronic decays, there should be 1 or 3 tracks inside the signal cone.

\begin{figure*}[t]
\captionsetup[subfigure]{labelformat=empty}
\centering
\begin{subfigure}{0.45\textwidth}
  \includegraphics[width=\textwidth]{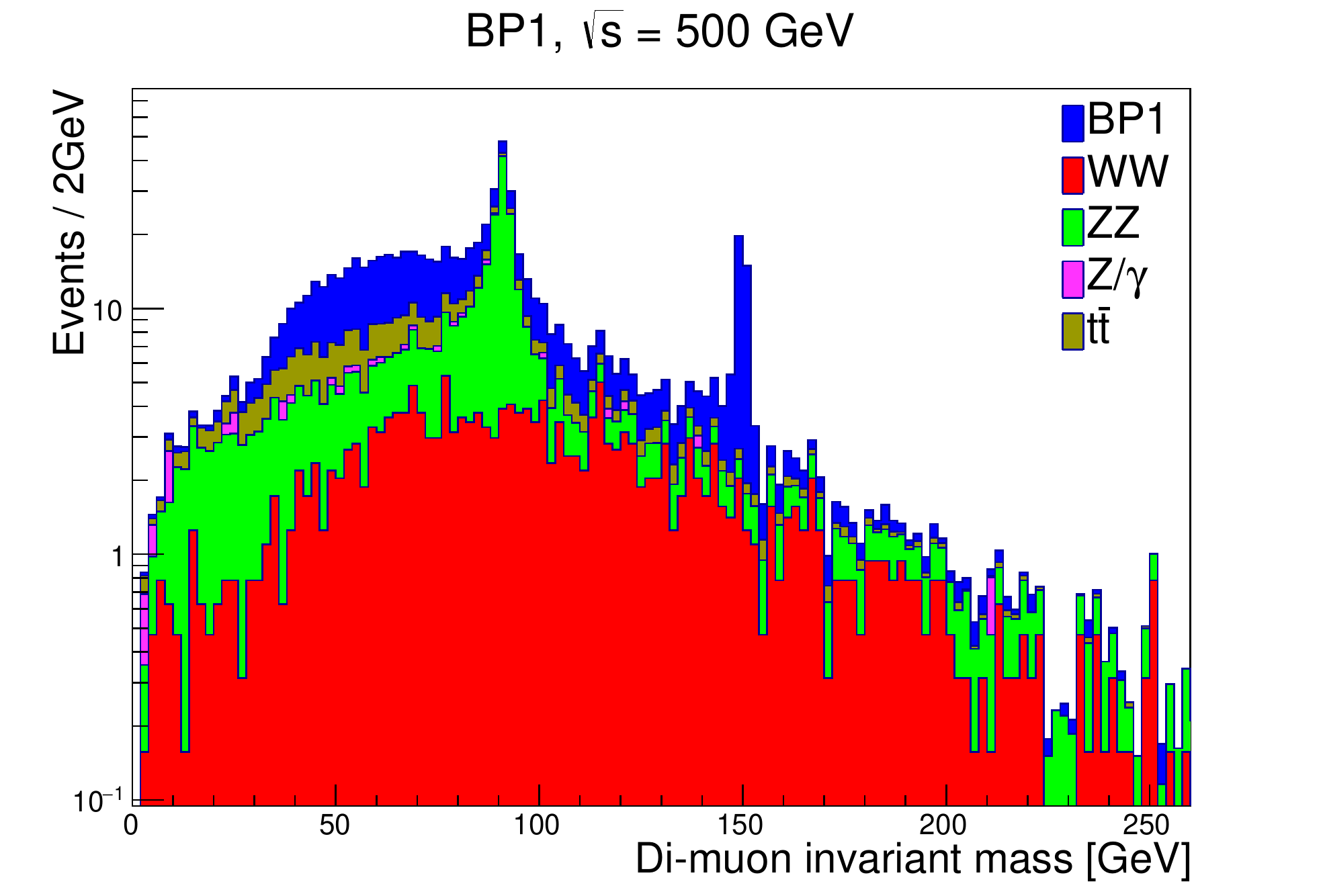}
  \caption{}
  \label{bp1}
\end{subfigure}%
\,\,
\begin{subfigure}{0.45\textwidth}
  \includegraphics[width=\textwidth]{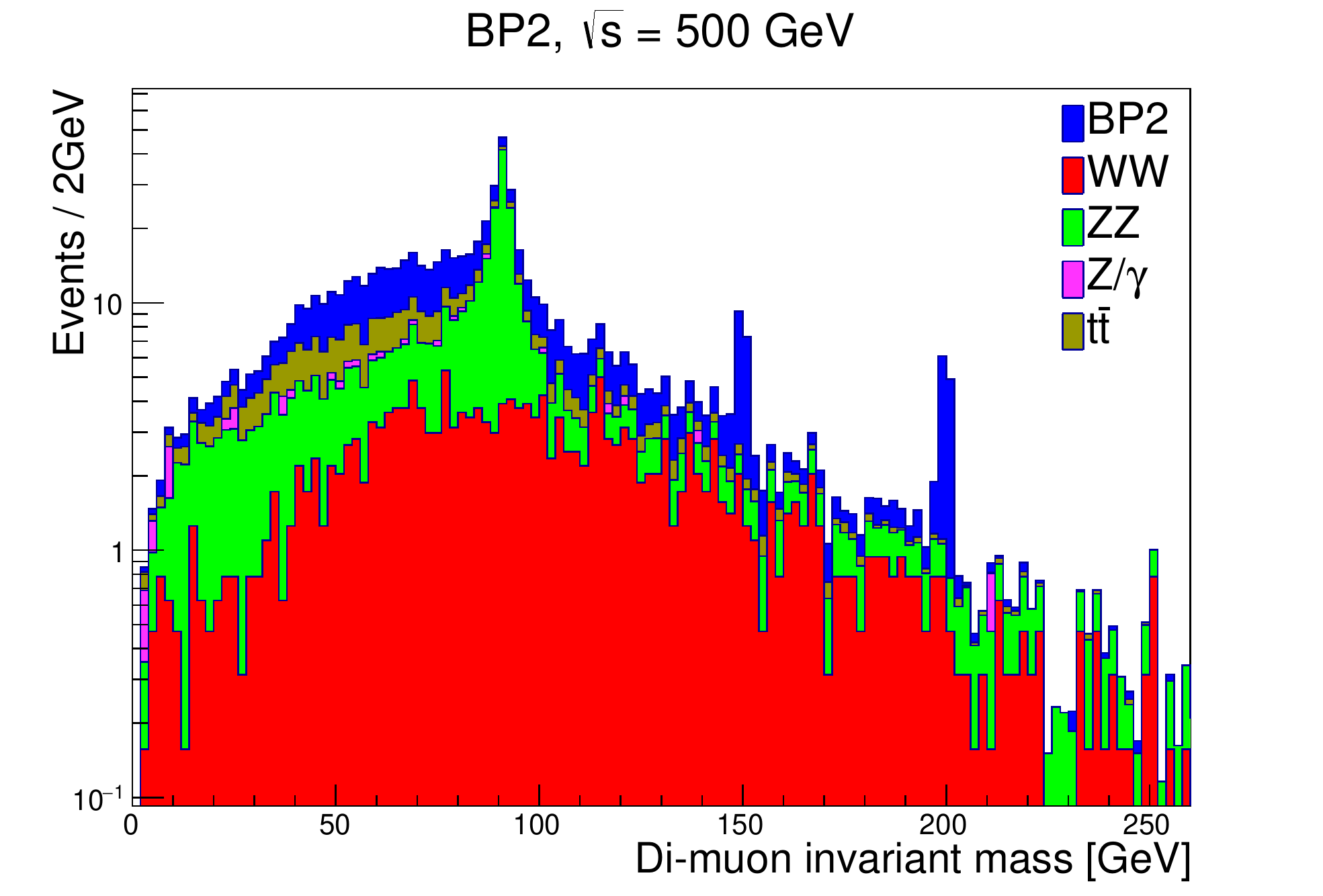}
  \caption{}
  \label{bp2}
\end{subfigure}%
\,\,
\begin{subfigure}{0.45\textwidth}
  \includegraphics[width=\textwidth]{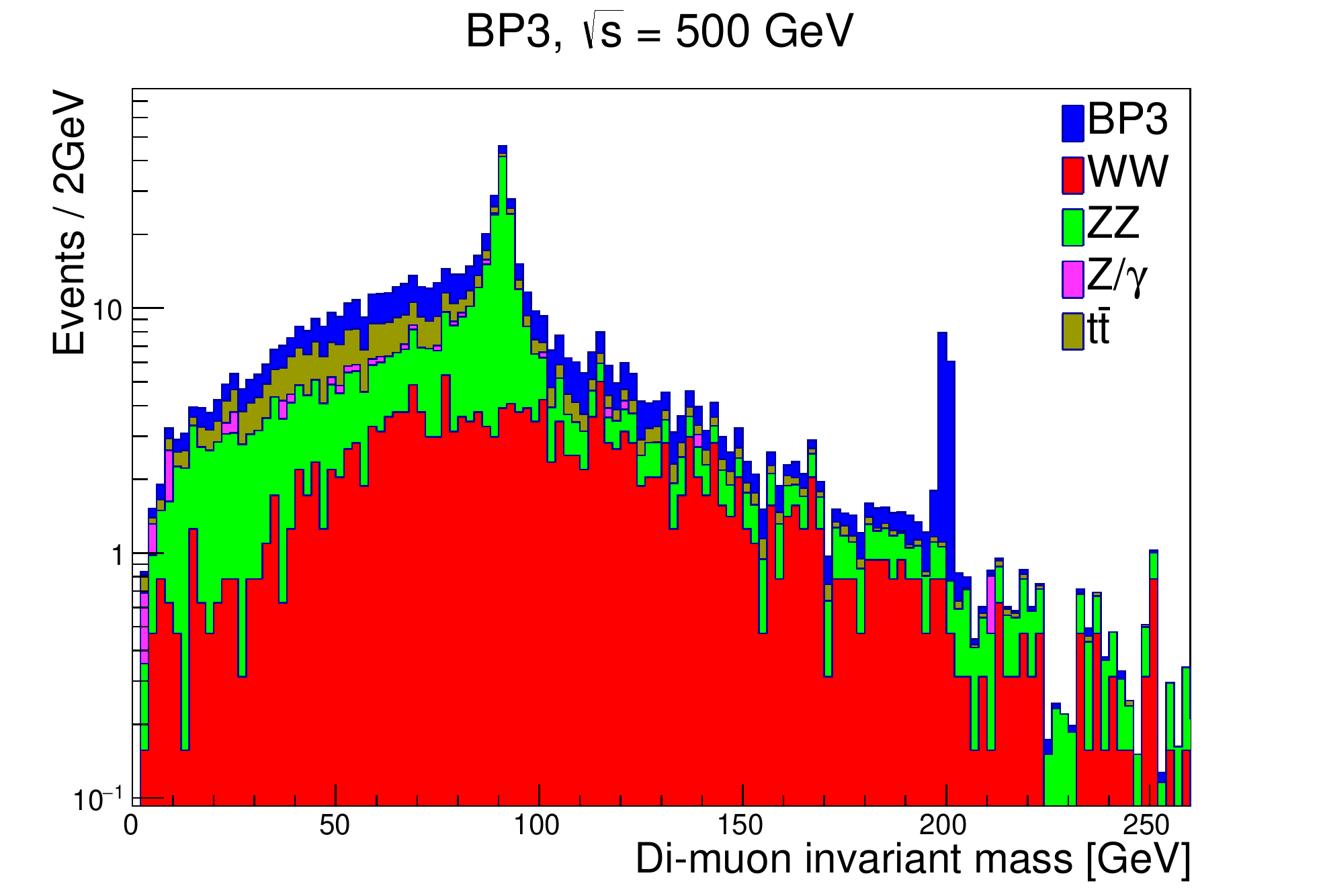}
  \caption{}
  \label{bp3}
\end{subfigure}%
\,\,
\begin{subfigure}{0.45\textwidth}
  \includegraphics[width=\textwidth]{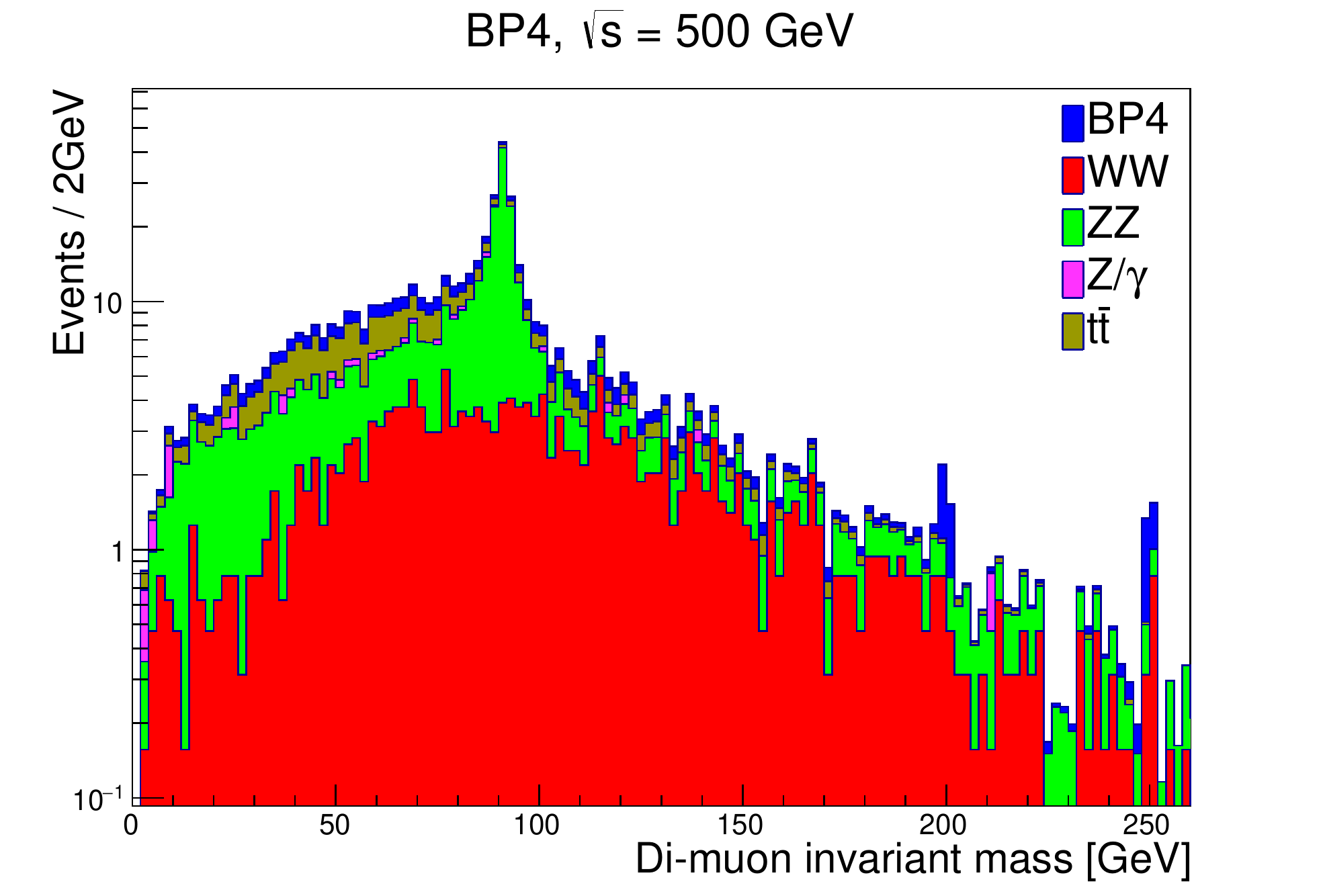}
  \caption{}
  \label{bp4}
\end{subfigure}%
\caption{The di-muon invariant mass distributions for different benchmark points at $\sqrt{s}=0.5$ TeV.}
\label{bp500}
\end{figure*}

\begin{figure*}[t]
\captionsetup[subfigure]{labelformat=empty}
\centering
\begin{subfigure}{0.45\textwidth}
  \includegraphics[width=\textwidth]{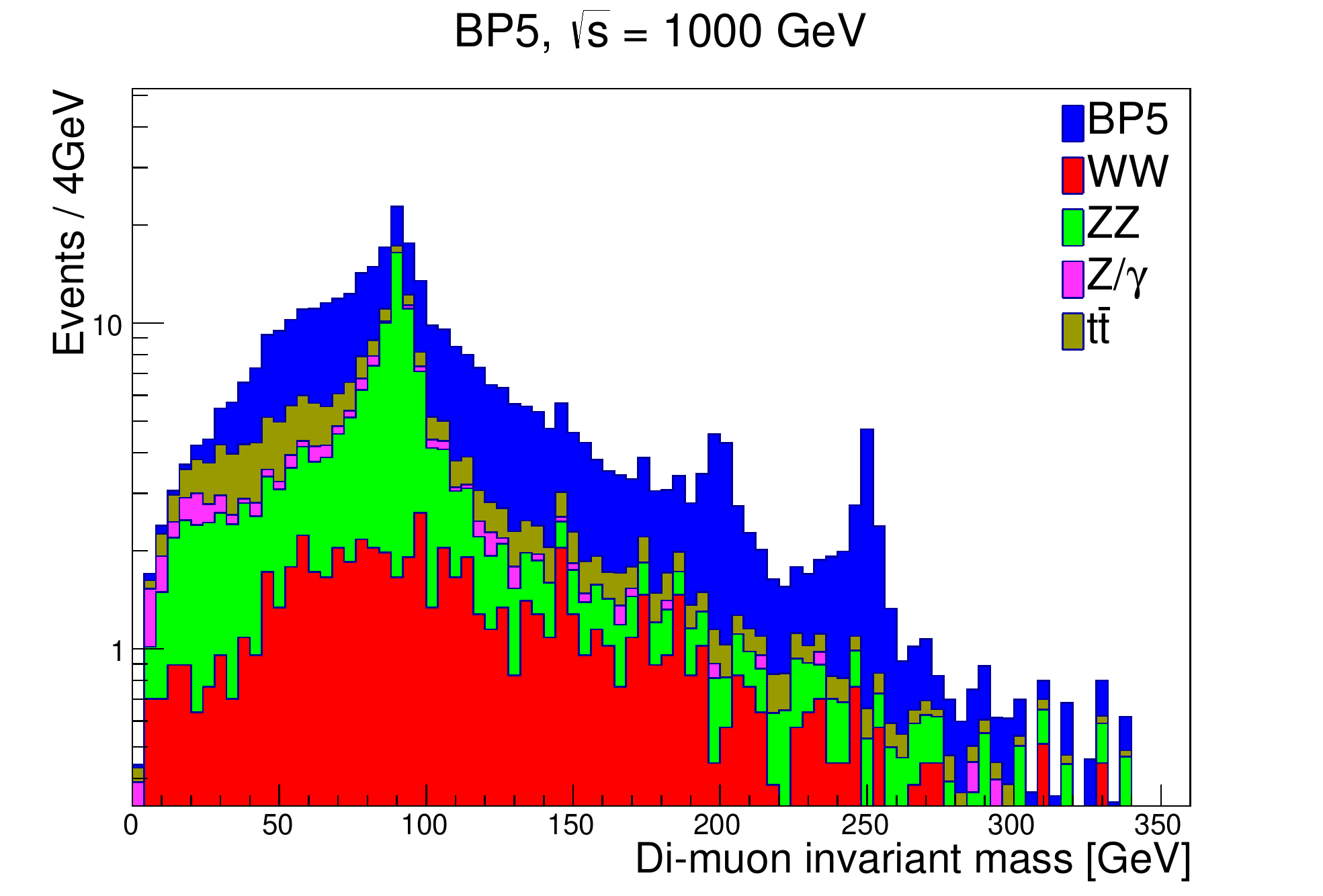}
  \caption{}
  \label{bp5}
\end{subfigure}%
\,\,
\begin{subfigure}{0.45\textwidth}
  \includegraphics[width=\textwidth]{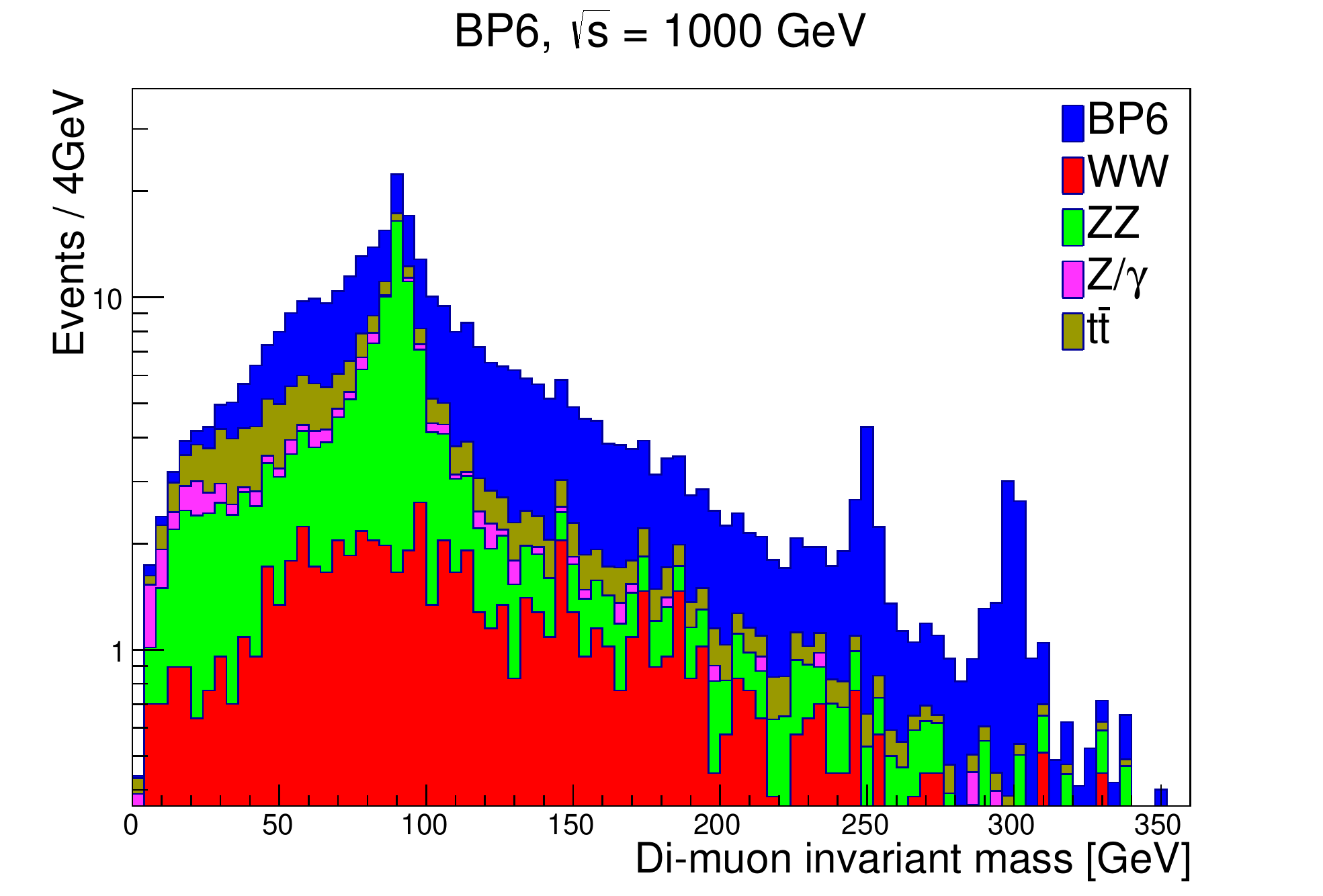}
  \caption{}
  \label{bp6}
\end{subfigure}%
\,\,
\begin{subfigure}{0.45\textwidth}
  \includegraphics[width=\textwidth]{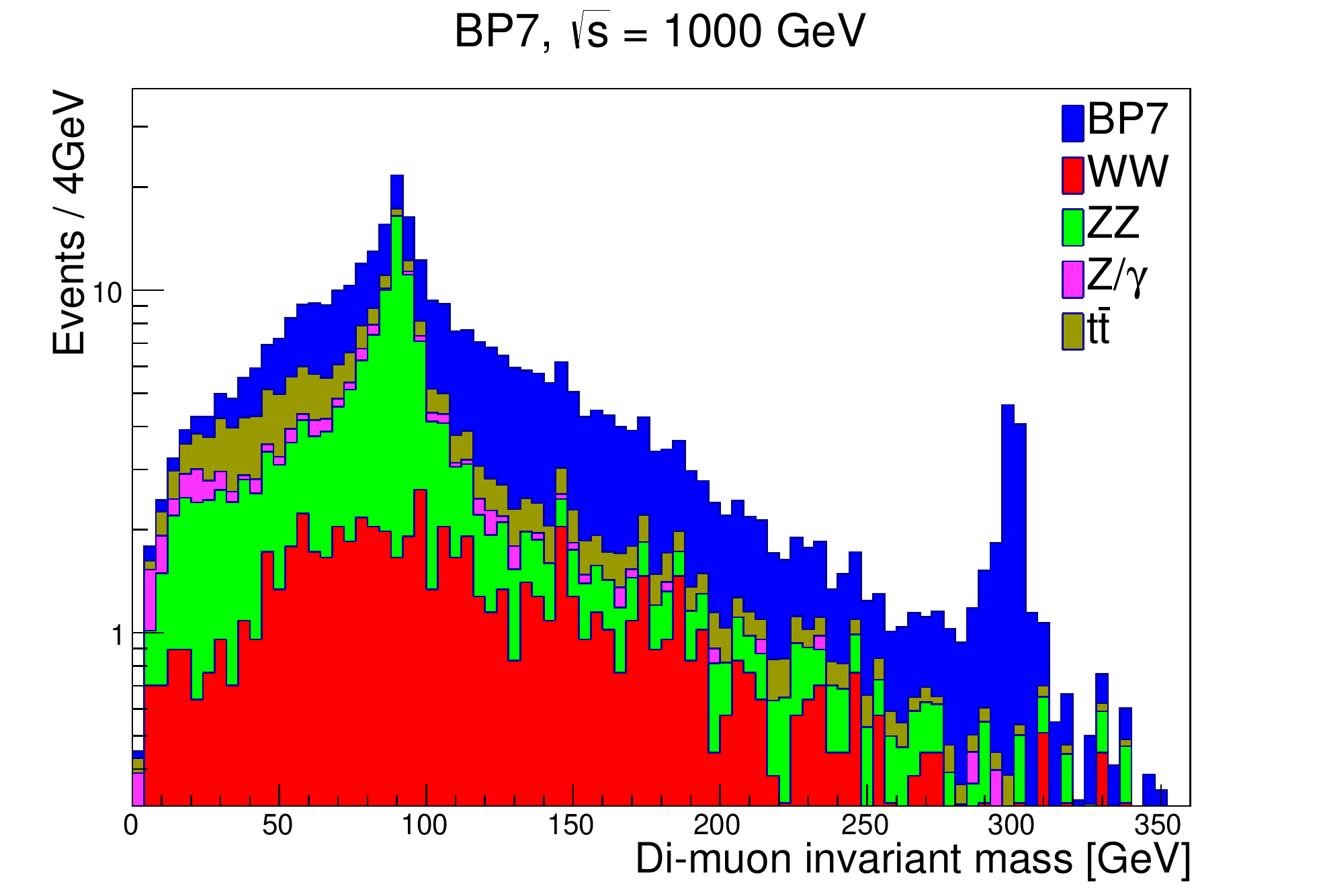}
  \caption{}
  \label{bp7}
\end{subfigure}%
\,\,
\begin{subfigure}{0.45\textwidth}
  \includegraphics[width=\textwidth]{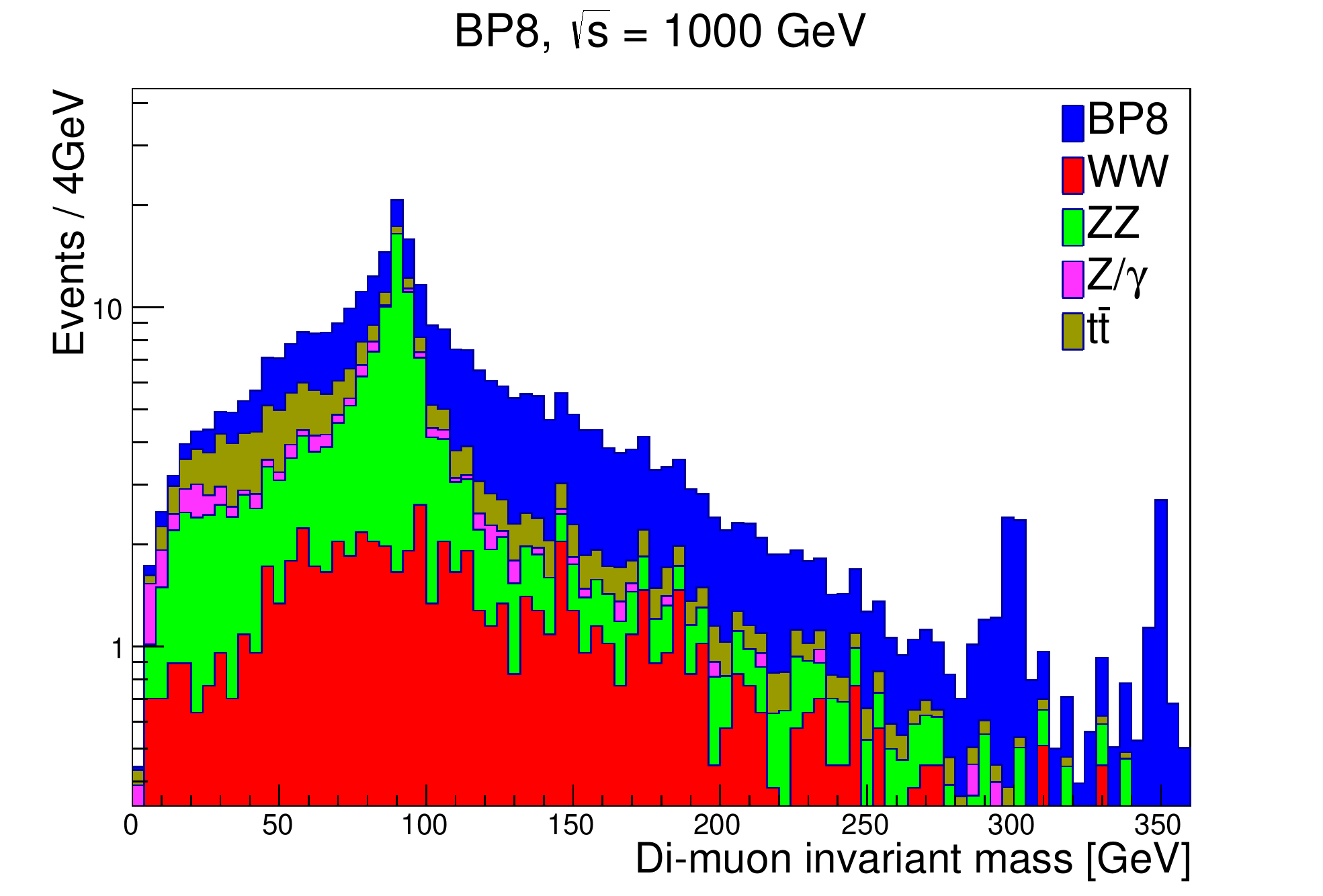}
  \caption{}
  \label{bp8}
\end{subfigure}%
\caption{The di-muon invariant mass distributions for different benchmark points at $\sqrt{s}=1$ TeV.}
\label{bp1000}
\end{figure*}

The muon selection is based on finding a muon satisfying the kinematic requirement as in \cref{kin2}. The muon track momentum resolution is also applied as $\sigma_{p_T}/p_{T}^2=2.10^{-5}$ GeV$^{-1}$ \cite{cliccdr}. An event has to have at least two jets identified as $\tau$ jets, at least two muons and missing transverse energy higher than 20 GeV to be selected. The last requirement is due to the fact that $\tau$ jets produce a source of neutrino (missing transverse energy) when they decay hadronically. This cut is useful for Drell-Yan background suppression.
\begin{equation}
E_{T}^{\textnormal{muon}}>10~\textnormal{GeV},~ |\eta|<1.5
\label{kin2}
\end{equation}
Events which pass all the above requirements are used for a di-muon invariant mass calculation whose distributions are shown in \cref{bp500} with $\sqrt{s}=0.5$ TeV and \cref{bp1000} with $\sqrt{s}=1$ TeV. As seen from \cref{bp500} and \cref{bp1000}, the signal is observable on top of the background in all benchmark points. The signal peaks are well separated from the $Z$ boson peak. In cases which involve different masses for $H$ and $A$, both Higgs bosons are observable. One may consider this process a promising way for a synchronized reconstruction of both scalar and pseudoscalar Higgs bosons. The signal and background selection efficiencies are calculated for each benchmark point as outlined in \cref{seleffs}.  The BP 4 (significance 1.6$\sigma$ at 0.5 TeV) receives a significance of 6.2$\sigma$ at 1 TeV. Therefore all benchmark points are observable at integrated luminosity of 1000 $fb^{-1}$.
\begin{table*}[h]
\centering
\begin{tabular}{|c||c|c||c|c|c|c||c|c||c|c|}
\hline
\multicolumn{10}{|c|}{$\sqrt{s}=500$ GeV}\\
BP & Signal &WW & ZZ & $Z/\gamma^*$ & $t\bar{t}$ & $S$ & $B$ & $S/B$ & $S/\sqrt{B}$ \\
\hline
1 & 0.01& \multirow{4}{*}{$3.10^{-5}$} & \multirow{4}{*}{$5.10^{-4}$} & \multirow{4}{*}{$6.10^{-7}$} & \multirow{4}{*}{$1.10^{-4}$} & 33 & 6.8 & 4.9 & 13 \\ \cline{1-2} \cline{7-10}
2 & 0.011& & & & & 23 & 9.8 & 2.3 & 7.4\\ \cline{1-2} \cline{7-10}
3 & 0.012& & & & & 13 & 3 & 4.2 & 7.3\\ \cline{1-2} \cline{7-10}
4 & 0.012& & & & & 3.4 & 4.7 & 0.7 & 1.6\\ \cline{1-2} \cline{7-10}
\hline
\multicolumn{10}{|c|}{$\sqrt{s}=1000$ GeV}\\
BP & Signal &WW & ZZ & $Z/\gamma^*$ & $t\bar{t}$ & $S$ & $B$ & $S/B$ & $S/\sqrt{B}$ \\
\hline
5 & 0.02 & \multirow{4}{*}{$3.10^{-5}$} & \multirow{4}{*}{$5.10^{-4}$} & \multirow{4}{*}{$2.10^{-6}$} & \multirow{4}{*}{$2.10^{-4}$} & 15 & 6.2 & 2.5 & 6.2\\ \cline{1-2} \cline{7-10}
6 & 0.02 & & & & & 12 & 4 & 3 & 6\\ \cline{1-2} \cline{7-10}
7 & 0.02 & & & & & 9.1 & 1.4 & 6.7 & 7.8\\ \cline{1-2} \cline{7-10}
8 & 0.03 & & & & &8.3 & 2 & 4.1 & 5.9 \\ \cline{1-2} \cline{7-10}
\hline
\end{tabular}
\caption{Signal and background selection efficiencies and final number of signal and background after the mass window cuts. The  $S/B$ ratio and the signal significance are also shown for an integrated luminosity of 1000 $fb^{-1}$. }
\label{seleffs}
\end{table*}
\section{Conclusions}
Signals of a neutral Higgs decay to leptons were analyzed at a linear collider operating at $\sqrt{s}=0.5$ and 1 TeV. The theoretical framework, i.e., 2HDM type IV, allows for enhancement of the leptonic decays of $H$ and $A$ bosons. Taking $e^+e^- \to AH$ as the signal, two pairs of di-muon and di-$\tau$ are produced. These events are similar in type with SM background processes like $ZZ$. However, a discriminating key feature, i.e., the invariant mass distributions of the two muons from the Higgs boson can be well separated from the Z pole mass if $m_H>m_Z$. With a detailed analysis of signal and background events, it was shown that the di-muon distribution from signal events can in fact be distinguished from the SM background. The signal to background ratio is large in all cases. The signal significance exceeds $5\sigma$ in all benchmark points at integrated luminosity of 1000 $fb^{-1}$.

\acknowledgements
I would like to thank Dr. Mogharrab for his efforts in providing the operational computing cluster of college of sciences at Shiraz University.
%


\begin{thebibliography}{29}%
\makeatletter
\providecommand \@ifxundefined [1]{%
 \@ifx{#1\undefined}
}%
\providecommand \@ifnum [1]{%
 \ifnum #1\expandafter \@firstoftwo
 \else \expandafter \@secondoftwo
 \fi
}%
\providecommand \@ifx [1]{%
 \ifx #1\expandafter \@firstoftwo
 \else \expandafter \@secondoftwo
 \fi
}%
\providecommand \natexlab [1]{#1}%
\providecommand \enquote  [1]{``#1''}%
\providecommand \bibnamefont  [1]{#1}%
\providecommand \bibfnamefont [1]{#1}%
\providecommand \citenamefont [1]{#1}%
\providecommand \href@noop [0]{\@secondoftwo}%
\providecommand \href [0]{\begingroup \@sanitize@url \@href}%
\providecommand \@href[1]{\@@startlink{#1}\@@href}%
\providecommand \@@href[1]{\endgroup#1\@@endlink}%
\providecommand \@sanitize@url [0]{\catcode `\\12\catcode `\$12\catcode
  `\&12\catcode `\#12\catcode `\^12\catcode `\_12\catcode `\%12\relax}%
\providecommand \@@startlink[1]{}%
\providecommand \@@endlink[0]{}%
\providecommand \url  [0]{\begingroup\@sanitize@url \@url }%
\providecommand \@url [1]{\endgroup\@href {#1}{\urlprefix }}%
\providecommand \urlprefix  [0]{URL }%
\providecommand \Eprint [0]{\href }%
\providecommand \doibase [0]{http://dx.doi.org/}%
\providecommand \selectlanguage [0]{\@gobble}%
\providecommand \bibinfo  [0]{\@secondoftwo}%
\providecommand \bibfield  [0]{\@secondoftwo}%
\providecommand \translation [1]{[#1]}%
\providecommand \BibitemOpen [0]{}%
\providecommand \bibitemStop [0]{}%
\providecommand \bibitemNoStop [0]{.\EOS\space}%
\providecommand \EOS [0]{\spacefactor3000\relax}%
\providecommand \BibitemShut  [1]{\csname bibitem#1\endcsname}%
\let\auto@bib@innerbib\@empty
\bibitem [{\citenamefont {Englert}\ and\ \citenamefont
  {Brout}(1964)}]{Englert1}%
  \BibitemOpen
  \bibfield  {author} {\bibinfo {author} {\bibfnamefont {F.}~\bibnamefont
  {Englert}}\ and\ \bibinfo {author} {\bibfnamefont {R.}~\bibnamefont
  {Brout}},\ }\href {\doibase 10.1103/PhysRevLett.13.321} {\bibfield  {journal}
  {\bibinfo  {journal} {Phys. Rev. Lett.}\ }\textbf {\bibinfo {volume} {13}},\
  \bibinfo {pages} {321} (\bibinfo {year} {1964})}\BibitemShut {NoStop}%
\bibitem [{\citenamefont {Higgs}(1964{\natexlab{a}})}]{Higgs1}%
  \BibitemOpen
  \bibfield  {author} {\bibinfo {author} {\bibfnamefont {P.~W.}\ \bibnamefont
  {Higgs}},\ }\href {\doibase 10.1103/PhysRevLett.13.508} {\bibfield  {journal}
  {\bibinfo  {journal} {Phys. Rev. Lett.}\ }\textbf {\bibinfo {volume} {13}},\
  \bibinfo {pages} {508} (\bibinfo {year} {1964}{\natexlab{a}})}\BibitemShut
  {NoStop}%
\bibitem [{\citenamefont {Higgs}(1964{\natexlab{b}})}]{Higgs2}%
  \BibitemOpen
  \bibfield  {author} {\bibinfo {author} {\bibfnamefont {P.~W.}\ \bibnamefont
  {Higgs}},\ }\href {\doibase 10.1016/0031-9163(64)91136-9} {\bibfield
  {journal} {\bibinfo  {journal} {Phys. Lett.}\ }\textbf {\bibinfo {volume}
  {12}},\ \bibinfo {pages} {132} (\bibinfo {year}
  {1964}{\natexlab{b}})}\BibitemShut {NoStop}%
\bibitem [{\citenamefont {Guralnik}\ \emph {et~al.}(1964)\citenamefont
  {Guralnik}, \citenamefont {Hagen},\ and\ \citenamefont {Kibble}}]{Kibble1}%
  \BibitemOpen
  \bibfield  {author} {\bibinfo {author} {\bibfnamefont {G.~S.}\ \bibnamefont
  {Guralnik}}, \bibinfo {author} {\bibfnamefont {C.~R.}\ \bibnamefont {Hagen}},
  \ and\ \bibinfo {author} {\bibfnamefont {T.~W.~B.}\ \bibnamefont {Kibble}},\
  }\href {\doibase 10.1103/PhysRevLett.13.585} {\bibfield  {journal} {\bibinfo
  {journal} {Phys. Rev. Lett.}\ }\textbf {\bibinfo {volume} {13}},\ \bibinfo
  {pages} {585} (\bibinfo {year} {1964})}\BibitemShut {NoStop}%
\bibitem [{\citenamefont {Higgs}(1966)}]{Higgs3}%
  \BibitemOpen
  \bibfield  {author} {\bibinfo {author} {\bibfnamefont {P.~W.}\ \bibnamefont
  {Higgs}},\ }\href {\doibase 10.1103/PhysRev.145.1156} {\bibfield  {journal}
  {\bibinfo  {journal} {Phys. Rev.}\ }\textbf {\bibinfo {volume} {145}},\
  \bibinfo {pages} {1156} (\bibinfo {year} {1966})}\BibitemShut {NoStop}%
\bibitem [{\citenamefont {Kibble}(1967)}]{Kibble2}%
  \BibitemOpen
  \bibfield  {author} {\bibinfo {author} {\bibfnamefont {T.~W.~B.}\
  \bibnamefont {Kibble}},\ }\href {\doibase 10.1103/PhysRev.155.1554}
  {\bibfield  {journal} {\bibinfo  {journal} {Phys. Rev.}\ }\textbf {\bibinfo
  {volume} {155}},\ \bibinfo {pages} {1554} (\bibinfo {year}
  {1967})}\BibitemShut {NoStop}%
\bibitem [{\citenamefont {Chatrchyan}\ \emph {et~al.}(2012)\citenamefont
  {Chatrchyan} \emph {et~al.}}]{HiggsObservationCMS}%
  \BibitemOpen
  \bibfield  {author} {\bibinfo {author} {\bibfnamefont {S.}~\bibnamefont
  {Chatrchyan}} \emph {et~al.} (\bibinfo {collaboration} {CMS}),\ }\href
  {\doibase 10.1016/j.physletb.2012.08.021} {\bibfield  {journal} {\bibinfo
  {journal} {Phys. Lett.}\ }\textbf {\bibinfo {volume} {B716}},\ \bibinfo
  {pages} {30} (\bibinfo {year} {2012})},\ \Eprint
  {http://arxiv.org/abs/1207.7235} {arXiv:1207.7235 [hep-ex]} \BibitemShut
  {NoStop}%
\bibitem [{\citenamefont {Aad}\ \emph {et~al.}(2012)\citenamefont {Aad} \emph
  {et~al.}}]{HiggsObservationATLAS}%
  \BibitemOpen
  \bibfield  {author} {\bibinfo {author} {\bibfnamefont {G.}~\bibnamefont
  {Aad}} \emph {et~al.} (\bibinfo {collaboration} {ATLAS}),\ }\href {\doibase
  10.1016/j.physletb.2012.08.020} {\bibfield  {journal} {\bibinfo  {journal}
  {Phys. Lett.}\ }\textbf {\bibinfo {volume} {B716}},\ \bibinfo {pages} {1}
  (\bibinfo {year} {2012})},\ \Eprint {http://arxiv.org/abs/1207.7214}
  {arXiv:1207.7214 [hep-ex]} \BibitemShut {NoStop}%
\bibitem [{\citenamefont {Lee}(1973)}]{2hdm1}%
  \BibitemOpen
  \bibfield  {author} {\bibinfo {author} {\bibfnamefont {T.~D.}\ \bibnamefont
  {Lee}},\ }\href {\doibase 10.1103/PhysRevD.8.1226} {\bibfield  {journal}
  {\bibinfo  {journal} {Phys. Rev.}\ }\textbf {\bibinfo {volume} {D8}},\
  \bibinfo {pages} {1226} (\bibinfo {year} {1973})}\BibitemShut {NoStop}%
\bibitem [{\citenamefont {Glashow}\ and\ \citenamefont
  {Weinberg}(1977)}]{2hdm2}%
  \BibitemOpen
  \bibfield  {author} {\bibinfo {author} {\bibfnamefont {S.~L.}\ \bibnamefont
  {Glashow}}\ and\ \bibinfo {author} {\bibfnamefont {S.}~\bibnamefont
  {Weinberg}},\ }\href {\doibase 10.1103/PhysRevD.15.1958} {\bibfield
  {journal} {\bibinfo  {journal} {Phys. Rev.}\ }\textbf {\bibinfo {volume}
  {D15}},\ \bibinfo {pages} {1958} (\bibinfo {year} {1977})}\BibitemShut
  {NoStop}%
\bibitem [{\citenamefont {Branco}(1980)}]{2hdm3}%
  \BibitemOpen
  \bibfield  {author} {\bibinfo {author} {\bibfnamefont {G.~C.}\ \bibnamefont
  {Branco}},\ }\href {\doibase 10.1103/PhysRevD.22.2901} {\bibfield  {journal}
  {\bibinfo  {journal} {Phys. Rev.}\ }\textbf {\bibinfo {volume} {D22}},\
  \bibinfo {pages} {2901} (\bibinfo {year} {1980})}\BibitemShut {NoStop}%
\bibitem [{\citenamefont {Aitchison}(2005)}]{MSSM1}%
  \BibitemOpen
  \bibfield  {author} {\bibinfo {author} {\bibfnamefont {I.~J.~R.}\
  \bibnamefont {Aitchison}},\ }\href@noop {} {\  (\bibinfo {year} {2005})},\
  \Eprint {http://arxiv.org/abs/hep-ph/0505105} {arXiv:hep-ph/0505105 [hep-ph]}
  \BibitemShut {NoStop}%
\bibitem [{\citenamefont {Ma}\ and\ \citenamefont {Ng}(1994)}]{MSSM2}%
  \BibitemOpen
  \bibfield  {author} {\bibinfo {author} {\bibfnamefont {E.}~\bibnamefont
  {Ma}}\ and\ \bibinfo {author} {\bibfnamefont {D.}~\bibnamefont {Ng}},\ }\href
  {\doibase 10.1103/PhysRevD.49.6164} {\bibfield  {journal} {\bibinfo
  {journal} {Phys. Rev.}\ }\textbf {\bibinfo {volume} {D49}},\ \bibinfo {pages}
  {6164} (\bibinfo {year} {1994})},\ \Eprint
  {http://arxiv.org/abs/hep-ph/9305230} {arXiv:hep-ph/9305230 [hep-ph]}
  \BibitemShut {NoStop}%
\bibitem [{\citenamefont {Djouadi}(2008)}]{MSSM3}%
  \BibitemOpen
  \bibfield  {author} {\bibinfo {author} {\bibfnamefont {A.}~\bibnamefont
  {Djouadi}},\ }\href {\doibase 10.1016/j.physrep.2007.10.005} {\bibfield
  {journal} {\bibinfo  {journal} {Phys. Rept.}\ }\textbf {\bibinfo {volume}
  {459}},\ \bibinfo {pages} {1} (\bibinfo {year} {2008})},\ \Eprint
  {http://arxiv.org/abs/hep-ph/0503173} {arXiv:hep-ph/0503173 [hep-ph]}
  \BibitemShut {NoStop}%
\bibitem [{\citenamefont {Mrazek}\ \emph {et~al.}(2011)\citenamefont {Mrazek},
  \citenamefont {Pomarol}, \citenamefont {Rattazzi}, \citenamefont {Redi},
  \citenamefont {Serra},\ and\ \citenamefont {Wulzer}}]{2hdm4_CompositeHiggs}%
  \BibitemOpen
  \bibfield  {author} {\bibinfo {author} {\bibfnamefont {J.}~\bibnamefont
  {Mrazek}}, \bibinfo {author} {\bibfnamefont {A.}~\bibnamefont {Pomarol}},
  \bibinfo {author} {\bibfnamefont {R.}~\bibnamefont {Rattazzi}}, \bibinfo
  {author} {\bibfnamefont {M.}~\bibnamefont {Redi}}, \bibinfo {author}
  {\bibfnamefont {J.}~\bibnamefont {Serra}}, \ and\ \bibinfo {author}
  {\bibfnamefont {A.}~\bibnamefont {Wulzer}},\ }\href {\doibase
  10.1016/j.nuclphysb.2011.07.008} {\bibfield  {journal} {\bibinfo  {journal}
  {Nucl. Phys.}\ }\textbf {\bibinfo {volume} {B853}},\ \bibinfo {pages} {1}
  (\bibinfo {year} {2011})},\ \Eprint {http://arxiv.org/abs/1105.5403}
  {arXiv:1105.5403 [hep-ph]} \BibitemShut {NoStop}%
\bibitem [{\citenamefont {Haber}\ and\ \citenamefont
  {O'Neil}(2006)}]{tanbsignificance}%
  \BibitemOpen
  \bibfield  {author} {\bibinfo {author} {\bibfnamefont {H.~E.}\ \bibnamefont
  {Haber}}\ and\ \bibinfo {author} {\bibfnamefont {D.}~\bibnamefont {O'Neil}},\
  }\href {\doibase 10.1103/PhysRevD.74.015018, 10.1103/PhysRevD.74.059905}
  {\bibfield  {journal} {\bibinfo  {journal} {Phys. Rev.}\ }\textbf {\bibinfo
  {volume} {D74}},\ \bibinfo {pages} {015018} (\bibinfo {year} {2006})},\
  \bibinfo {note} {[Erratum: Phys. Rev.D74,no.5,059905(2006)]},\ \Eprint
  {http://arxiv.org/abs/hep-ph/0602242} {arXiv:hep-ph/0602242 [hep-ph]}
  \BibitemShut {NoStop}%
\bibitem [{\citenamefont {Branco}\ \emph {et~al.}(2012)\citenamefont {Branco},
  \citenamefont {Ferreira}, \citenamefont {Lavoura}, \citenamefont {Rebelo},
  \citenamefont {Sher},\ and\ \citenamefont {Silva}}]{2hdm_TheoryPheno}%
  \BibitemOpen
  \bibfield  {author} {\bibinfo {author} {\bibfnamefont {G.~C.}\ \bibnamefont
  {Branco}}, \bibinfo {author} {\bibfnamefont {P.~M.}\ \bibnamefont
  {Ferreira}}, \bibinfo {author} {\bibfnamefont {L.}~\bibnamefont {Lavoura}},
  \bibinfo {author} {\bibfnamefont {M.~N.}\ \bibnamefont {Rebelo}}, \bibinfo
  {author} {\bibfnamefont {M.}~\bibnamefont {Sher}}, \ and\ \bibinfo {author}
  {\bibfnamefont {J.~P.}\ \bibnamefont {Silva}},\ }\href {\doibase
  10.1016/j.physrep.2012.02.002} {\bibfield  {journal} {\bibinfo  {journal}
  {Phys. Rept.}\ }\textbf {\bibinfo {volume} {516}},\ \bibinfo {pages} {1}
  (\bibinfo {year} {2012})},\ \Eprint {http://arxiv.org/abs/1106.0034}
  {arXiv:1106.0034 [hep-ph]} \BibitemShut {NoStop}%
\bibitem [{\citenamefont {Mahmoudi}\ and\ \citenamefont
  {Stal}(2010)}]{FMahmoudi}%
  \BibitemOpen
  \bibfield  {author} {\bibinfo {author} {\bibfnamefont {F.}~\bibnamefont
  {Mahmoudi}}\ and\ \bibinfo {author} {\bibfnamefont {O.}~\bibnamefont
  {Stal}},\ }\href {\doibase 10.1103/PhysRevD.81.035016} {\bibfield  {journal}
  {\bibinfo  {journal} {Phys. Rev.}\ }\textbf {\bibinfo {volume} {D81}},\
  \bibinfo {pages} {035016} (\bibinfo {year} {2010})},\ \Eprint
  {http://arxiv.org/abs/0907.1791} {arXiv:0907.1791 [hep-ph]} \BibitemShut
  {NoStop}%
\bibitem [{\citenamefont {Kanemura}\ \emph {et~al.}(2012)\citenamefont
  {Kanemura}, \citenamefont {Tsumura},\ and\ \citenamefont
  {Yokoya}}]{kanemura}%
  \BibitemOpen
  \bibfield  {author} {\bibinfo {author} {\bibfnamefont {S.}~\bibnamefont
  {Kanemura}}, \bibinfo {author} {\bibfnamefont {K.}~\bibnamefont {Tsumura}}, \
  and\ \bibinfo {author} {\bibfnamefont {H.}~\bibnamefont {Yokoya}},\ }\href
  {\doibase 10.1103/PhysRevD.85.095001} {\bibfield  {journal} {\bibinfo
  {journal} {Phys. Rev.}\ }\textbf {\bibinfo {volume} {D85}},\ \bibinfo {pages}
  {095001} (\bibinfo {year} {2012})},\ \Eprint {http://arxiv.org/abs/1111.6089}
  {arXiv:1111.6089 [hep-ph]} \BibitemShut {NoStop}%
\bibitem [{\citenamefont {Davidson}\ and\ \citenamefont
  {Haber}(2005)}]{2hdm_HiggsSector1}%
  \BibitemOpen
  \bibfield  {author} {\bibinfo {author} {\bibfnamefont {S.}~\bibnamefont
  {Davidson}}\ and\ \bibinfo {author} {\bibfnamefont {H.~E.}\ \bibnamefont
  {Haber}},\ }\href {\doibase 10.1103/PhysRevD.72.099902,
  10.1103/PhysRevD.72.035004} {\bibfield  {journal} {\bibinfo  {journal} {Phys.
  Rev.}\ }\textbf {\bibinfo {volume} {D72}},\ \bibinfo {pages} {035004}
  (\bibinfo {year} {2005})},\ \bibinfo {note} {[Erratum: Phys.
  Rev.D72,099902(2005)]},\ \Eprint {http://arxiv.org/abs/hep-ph/0504050}
  {arXiv:hep-ph/0504050 [hep-ph]} \BibitemShut {NoStop}%
\bibitem [{\citenamefont {Barger}\ \emph {et~al.}(1990)\citenamefont {Barger},
  \citenamefont {Hewett},\ and\ \citenamefont {Phillips}}]{Barger_2hdmTypes}%
  \BibitemOpen
  \bibfield  {author} {\bibinfo {author} {\bibfnamefont {V.~D.}\ \bibnamefont
  {Barger}}, \bibinfo {author} {\bibfnamefont {J.~L.}\ \bibnamefont {Hewett}},
  \ and\ \bibinfo {author} {\bibfnamefont {R.~J.~N.}\ \bibnamefont
  {Phillips}},\ }\href {\doibase 10.1103/PhysRevD.41.3421} {\bibfield
  {journal} {\bibinfo  {journal} {Phys. Rev.}\ }\textbf {\bibinfo {volume}
  {D41}},\ \bibinfo {pages} {3421} (\bibinfo {year} {1990})}\BibitemShut
  {NoStop}%
\bibitem [{\citenamefont {Aoki}\ \emph {et~al.}(2009)\citenamefont {Aoki},
  \citenamefont {Kanemura}, \citenamefont {Tsumura},\ and\ \citenamefont
  {Yagyu}}]{2hdm_HiggsSector2}%
  \BibitemOpen
  \bibfield  {author} {\bibinfo {author} {\bibfnamefont {M.}~\bibnamefont
  {Aoki}}, \bibinfo {author} {\bibfnamefont {S.}~\bibnamefont {Kanemura}},
  \bibinfo {author} {\bibfnamefont {K.}~\bibnamefont {Tsumura}}, \ and\
  \bibinfo {author} {\bibfnamefont {K.}~\bibnamefont {Yagyu}},\ }\href
  {\doibase 10.1103/PhysRevD.80.015017} {\bibfield  {journal} {\bibinfo
  {journal} {Phys. Rev.}\ }\textbf {\bibinfo {volume} {D80}},\ \bibinfo {pages}
  {015017} (\bibinfo {year} {2009})},\ \Eprint {http://arxiv.org/abs/0902.4665}
  {arXiv:0902.4665 [hep-ph]} \BibitemShut {NoStop}%
\bibitem [{\citenamefont {Misiak}\ \emph {et~al.}(2015)\citenamefont {Misiak}
  \emph {et~al.}}]{Misiak}%
  \BibitemOpen
  \bibfield  {author} {\bibinfo {author} {\bibfnamefont {M.}~\bibnamefont
  {Misiak}} \emph {et~al.},\ }\href {\doibase 10.1103/PhysRevLett.114.221801}
  {\bibfield  {journal} {\bibinfo  {journal} {Phys. Rev. Lett.}\ }\textbf
  {\bibinfo {volume} {114}},\ \bibinfo {pages} {221801} (\bibinfo {year}
  {2015})},\ \Eprint {http://arxiv.org/abs/1503.01789} {arXiv:1503.01789
  [hep-ph]} \BibitemShut {NoStop}%
\bibitem [{\citenamefont {Dorsch}\ \emph {et~al.}(2014)\citenamefont {Dorsch},
  \citenamefont {Huber}, \citenamefont {Mimasu},\ and\ \citenamefont
  {No}}]{TypeI_LHC}%
  \BibitemOpen
  \bibfield  {author} {\bibinfo {author} {\bibfnamefont {G.~C.}\ \bibnamefont
  {Dorsch}}, \bibinfo {author} {\bibfnamefont {S.~J.}\ \bibnamefont {Huber}},
  \bibinfo {author} {\bibfnamefont {K.}~\bibnamefont {Mimasu}}, \ and\ \bibinfo
  {author} {\bibfnamefont {J.~M.}\ \bibnamefont {No}},\ }\href {\doibase
  10.1103/PhysRevLett.113.211802} {\bibfield  {journal} {\bibinfo  {journal}
  {Phys. Rev. Lett.}\ }\textbf {\bibinfo {volume} {113}},\ \bibinfo {pages}
  {211802} (\bibinfo {year} {2014})},\ \Eprint {http://arxiv.org/abs/1405.5537}
  {arXiv:1405.5537 [hep-ph]} \BibitemShut {NoStop}%
\bibitem [{\citenamefont {Grimus}\ \emph {et~al.}(2008)\citenamefont {Grimus},
  \citenamefont {Lavoura}, \citenamefont {Ogreid},\ and\ \citenamefont
  {Osland}}]{drho}%
  \BibitemOpen
  \bibfield  {author} {\bibinfo {author} {\bibfnamefont {W.}~\bibnamefont
  {Grimus}}, \bibinfo {author} {\bibfnamefont {L.}~\bibnamefont {Lavoura}},
  \bibinfo {author} {\bibfnamefont {O.~M.}\ \bibnamefont {Ogreid}}, \ and\
  \bibinfo {author} {\bibfnamefont {P.}~\bibnamefont {Osland}},\ }\href
  {\doibase 10.1088/0954-3899/35/7/075001} {\bibfield  {journal} {\bibinfo
  {journal} {J. Phys.}\ }\textbf {\bibinfo {volume} {G35}},\ \bibinfo {pages}
  {075001} (\bibinfo {year} {2008})},\ \Eprint {http://arxiv.org/abs/0711.4022}
  {arXiv:0711.4022 [hep-ph]} \BibitemShut {NoStop}%
\bibitem [{\citenamefont {Sjostrand}\ \emph {et~al.}(2008)\citenamefont
  {Sjostrand}, \citenamefont {Mrenna},\ and\ \citenamefont {Skands}}]{pythia}%
  \BibitemOpen
  \bibfield  {author} {\bibinfo {author} {\bibfnamefont {T.}~\bibnamefont
  {Sjostrand}}, \bibinfo {author} {\bibfnamefont {S.}~\bibnamefont {Mrenna}}, \
  and\ \bibinfo {author} {\bibfnamefont {P.~Z.}\ \bibnamefont {Skands}},\
  }\href {\doibase 10.1016/j.cpc.2008.01.036} {\bibfield  {journal} {\bibinfo
  {journal} {Comput. Phys. Commun.}\ }\textbf {\bibinfo {volume} {178}},\
  \bibinfo {pages} {852} (\bibinfo {year} {2008})},\ \Eprint
  {http://arxiv.org/abs/0710.3820} {arXiv:0710.3820 [hep-ph]} \BibitemShut
  {NoStop}%
\bibitem [{\citenamefont {Cacciari}(2006)}]{fastjet1}%
  \BibitemOpen
  \bibfield  {author} {\bibinfo {author} {\bibfnamefont {M.}~\bibnamefont
  {Cacciari}},\ }in\ \href@noop {} {\emph {\bibinfo {booktitle} {{Deep
  inelastic scattering. Proceedings, 14th International Workshop, DIS 2006,
  Tsukuba, Japan, April 20-24, 2006}}}}\ (\bibinfo {year} {2006})\ pp.\
  \bibinfo {pages} {487--490},\ \bibinfo {note} {[,125(2006)]},\ \Eprint
  {http://arxiv.org/abs/hep-ph/0607071} {arXiv:hep-ph/0607071 [hep-ph]}
  \BibitemShut {NoStop}%
\bibitem [{\citenamefont {Cacciari}\ \emph {et~al.}(2012)\citenamefont
  {Cacciari}, \citenamefont {Salam},\ and\ \citenamefont {Soyez}}]{fastjet2}%
  \BibitemOpen
  \bibfield  {author} {\bibinfo {author} {\bibfnamefont {M.}~\bibnamefont
  {Cacciari}}, \bibinfo {author} {\bibfnamefont {G.~P.}\ \bibnamefont {Salam}},
  \ and\ \bibinfo {author} {\bibfnamefont {G.}~\bibnamefont {Soyez}},\ }\href
  {\doibase 10.1140/epjc/s10052-012-1896-2} {\bibfield  {journal} {\bibinfo
  {journal} {Eur. Phys. J.}\ }\textbf {\bibinfo {volume} {C72}},\ \bibinfo
  {pages} {1896} (\bibinfo {year} {2012})},\ \Eprint
  {http://arxiv.org/abs/1111.6097} {arXiv:1111.6097 [hep-ph]} \BibitemShut
  {NoStop}%
\bibitem [{\citenamefont {Linssen}\ \emph {et~al.}(2012)\citenamefont
  {Linssen}, \citenamefont {Miyamoto}, \citenamefont {Stanitzki},\ and\
  \citenamefont {Weerts}}]{cliccdr}%
  \BibitemOpen
  \bibfield  {author} {\bibinfo {author} {\bibfnamefont {L.}~\bibnamefont
  {Linssen}}, \bibinfo {author} {\bibfnamefont {A.}~\bibnamefont {Miyamoto}},
  \bibinfo {author} {\bibfnamefont {M.}~\bibnamefont {Stanitzki}}, \ and\
  \bibinfo {author} {\bibfnamefont {H.}~\bibnamefont {Weerts}},\ }\href
  {\doibase 10.5170/CERN-2012-003} {\  (\bibinfo {year} {2012}),\
  10.5170/CERN-2012-003},\ \Eprint {http://arxiv.org/abs/1202.5940}
  {arXiv:1202.5940 [physics.ins-det]} \BibitemShut {NoStop}%
\end{thebibliography}

\end{document}